\documentclass[12pt,a4paper]{article}

\begin{document}

\begin{center}

{\Large\bf LanHEP - a package for automatic generation of Feynman rules from
 the Lagrangian. Updated version 3.1} \\[8mm]

{\large  A.~Semenov. }\\[4mm]

{\it  Joint Institute of Nuclear research, JINR, 141980 Dubna, Russia  }\\[4mm]

\end{center}

 \begin{abstract}
 We present a new version 3.1 of the LanHEP 
 software package. New features of the program include tools for the 
 models with extra dimensions,
 implementation of the particle classes for FeynArts output and
 using templates 
 with LanHEP statements.
 \end{abstract}

\section*{Introduction}
The LanHEP program \cite{lanhep30} is developed for Feynman
rules generation from the Lagrangian. It reads
the Lagrangian written in a compact form, close to the one used in 
publications. It means that Lagrangian terms can be written 
with summation over indices of broken symmetries and using special symbols
for complicated expressions, such as covariant derivative and  
strength tensor for gauge fields. Supersymmetric theories can be described 
using the superpotential
    formalism and the 2-component fermion notation. The output
is Feynman rules in terms of physical fields and independent parameters
 in the form of CompHEP \cite{comphep} or CalcHEP \cite{calchep}
model files, which allows one to start calculations of processes in
the new physical model. Alternatively, Feynman rules can be generated
in FeynArts \cite{FeynArts} format or as LaTeX table. The program can also generate 
one-loop counterterms in the FeynArts format.
This note describes new features of the version 3.1 of the LanHEP 
 package, including tools  for the 
 models with extra dimensions,
 implementation of the particle classes for FeynArts output and
 using templates with LanHEP statements.
 The package can be douwnoaded from 
 { \tt http://theory.sinp.msu.ru/\~{}semenov/lanhep.html}

\section{Models with extra dimensions}
 
A new feature in LanHEP helps to generate Kaluza-Klein modes for
particles in models with additional
dimensions. In case of the Minimal Universal Extra Dimension Model \cite{mued},
the photon field in 5 dimensions can be projected into 4-dimensional space
\begin{eqnarray*}
  A_{\mu}\left(x^{\mu},y\right) & = & \frac{1}{\sqrt{\pi R}}\left\{ A_{\mu}^{(0)}(x)+\sqrt{2}\sum_{n=1}^{\infty}A_{\mu}^{(n)}(x)\,\cos\left(\frac{ny}{R}\right)\right\},~ \\
  A_{5}\left(x^{\mu},y\right) & = & \sqrt{\frac{2}{\pi R}}\sum_{n=1}^{\infty}A_{5}^{(n)}(x)\,\sin\left(\frac{ny}{R}\right)
\end{eqnarray*}

LanHEP allows to 
expand the 4-dimensional field $A_\mu(x^\mu)$ into the sum 
$A^{(0)}_\mu(x)$ and
 Kaluza-Klein modes $A_\mu^{(n)}(x)$ as in the
right-hand part of the equation above, by the statement
\begin{quote} {\tt transform A -> A + (A1*cos(1) + A2*cos(2))*Sqrt2.}
\end{quote}
Here {\tt sin(\sl N\tt )}, {\tt cos(\sl N\tt)} correspond to KK-mode number $N$,
and  these function will be integrated out using orthogonality relation after
constructing the Lagrangian. We have introduced two KK modes in this example,
but one can write only one or add more. 
Similar prescription can be written for scalars and spinors.
The {\tt transform} statements allows to expand with KK modes particles
in the existing LanHEP model without modifying the statements which
describe Lagrangian terms.

 One also can define the scalar field corresponding to the fifth component
 of the photon:
\begin{quote} {\tt let A5 = (s1*sin(1) + s2*sin(2))*Sqrt2.}
\end{quote}
Here {\tt s1}, {\tt s2} are scalar fields which should be declared as particles
before. 
One can write the interactions of the 5th components of vector fields
explicitly, using symbols like {\tt A5} defined above and the symbol
{\tt deriv5} for $\partial_5$. Note that {\tt deriv5} differentiate only 
{\tt sin} and {\tt cos} functions, and multiply it by the mode number, 
so the scale factor should be written explicitly next to {\tt deriv5}. 

LanHEP can generate the interaction of 5th components
automatically, by adding the product of the 5th components 
to each convolution of vector indices. To do this, one should use the
{\tt ued\_5th} statement to define the 5th components of vectors:
\begin{quote} {\tt ued\_5th deriv -> deriv5/R, A -> A5.}
\end{quote}
Here R is the scale parameter. One also can specify the scale parameter
in the second argument of {\tt sin} or {\tt cos} function, like {\tt cos(1,invR)},
where invR is defined as 1/R. In this case, one should write {\tt deriv->deriv5}
in the {\tt ued\_5th} statement.

\section{Classes in FeynArts output}

FeynArts allows to combine particles with similar properties into {\sl classes}.
By default, LanHEP generates the model where each particle has its own class.
It is possible to combine several particles into a class by the {\tt class} 
statement. For example
\begin{quote} {\tt class lpc=[e,m,l].} \end{quote}
joins the electron, muon and tau-lepton into the class {\tt lpc} (charged
lepton). So, the vertices with these three particles will be joined to 
describe generic lepton interaction with other fields. This feature allows to 
decrease
the number of vertices and speeds up calculations. The particles in the 
class must have the same spin and color, however
it is possible to combine into a class  particles with different electric 
charge, or scalars that are CP-even and CP-odd scalars.

\section{Using templates}

A model description often includes several statements with the same
structure. For example, the declaration of the parameters which are 
elements of some mixing matrix, evaluated by external function reads
\begin{quote} {\tt parameter Zn11=MixMatr(neu,1,1), Zn12=MixMatr(neu,1,2), ...
} \end{quote}
where '...' stands for all other definitions for this matrix. The declaration
of these parameters can be written in a simplier form:
\begin{quote} {\tt  
   \_x=1-4, \_y=1-4 in parameter Zn\_x\_y=MixMatr(neu, \_x, \_y).}
\end{quote}
Here the {\tt parameter} statement will be invoked several times for all
possible combinations of values for symbols like \_x, making the substitution
when \_x appears into one of values, and creating names { \tt Zn11, Zn12, ...}
from template {\tt Zn\_x\_y}. These symbols must have one letter. The values 
can be set as {\tt \_x=1-4}  or {\tt \_x=[1,2,3,4]}.
The latter form is useful when substitution values are not sequential numbers,
for example values can be particle names.
The prefix with  keyword {\tt 'in'} can be applied to any LanHEP statement.

Another way to execute a statement several times with different names of
parameters is to use the keyword {\tt 'where'}. This feature was already 
used in earlier versions in the lagrangian terms. For example 
\begin{verbatim}
lterm  	anti(psi)*gamma*(1+g5)/2*(i*deriv - Y*g1*B1)*psi  where 
           psi=e, Y= -1;   psi=m, Y= -1;   psi=l, Y= -1;
           psi=u, Y=  2/3; psi=c, Y=  2/3; psi=t, Y=  2/3;
           psi=d, Y= -1/3; psi=s, Y= -1/3; psi=b, Y= -1/3.
\end{verbatim}
describes the gauge interaction for quarks and leptons, Y is hypercharge.
Now the substitution with the keywors {\tt 'where'} can be applied as postfix
to any LanHEP statement, and the description of substitutions can 
be made in a shorter form:
\begin{quote}
{\tt lterm ... where psi=[e,m,l,u,c,t,d,s,b], Y=[-1,-1,-1,2/3,2/3,2/3,-1/3,
-1/3,-1/3]}.
\end{quote} 
The lists of the values for names of substitutions must be the same length,
and this length is the number of times the statement is executed. At each 
execution the next values from the lists are used for substitution symbols.

When it is necessary to execute the statement with all combinations of 
substitutions in two (or more) list, one can use nested keywords {\tt 'where'}.
For example, deriving Yukawa terms from the superpotential may read
\begin{quote}{\tt  (lterm -df(superW,Ai,Aj)*Fi*Fj/2 + AddHermConj \\            
			where Ai=[h1,h2],Fi=[fh1,fh2] ) where Aj=[h1,h2],Fj=[fh1,fh2].}
\end{quote}

Both {\tt 'in'} and {\tt 'where'} keywords can be used if it is necessary 
to combine templates for names like {\tt Zn\_i\_j} and substitutions for group 
of symbols. The construction will look:
\begin{quote} 
{\tt \_x=name in \sl statement \tt where name= \sl values, ... .}
\end{quote}
In this example the keyword {\tt 'where'} will substitute {\tt name} with given
values, which then will be used by keyword {\tt 'in'} to substitute symbol 
{\tt '\_x'} in the {\sl statement}. If the keyword {\tt 'in'} is needed to make
substitutions before {\tt 'where'}, brackets can be used:
\begin{quote} 
{\tt \_x=[u,d,c,s,t,b] in ( \sl statement \tt where mass=M\_x, ...) .}
\end{quote}
In general, any construction with {\tt 'in'} or {\tt 'where'} can be included 
in brackets and next {\tt 'in'} or {\tt 'where'} appended. The outermost 
keyword makes its substitution first.

 \section*{Acknowledgments}
  This work was  supported in part by the GDRI-ACPP of CNRS and by 
  the ANR project {\tt ToolsDMColl}, BLAN07-2-194882.
This work  was  also supported by the Russian foundation for Basic Research, 
grant RFBR-08-02-92499-a.

 \end{document}